\documentclass[a4paper,12pt]{article}
\pdfoutput=1 % if your are submitting a pdflatex (i.e. if you have
\usepackage{graphicx}
\usepackage{jheppub}
\usepackage{ytableau}
\usepackage{pgf,pgfarrows,pgfnodes}
\usepackage{graphics}
\usepackage{hyperref}
\usepackage{tikz}
\usepackage{xparse}
\usepackage{subfigure}
\newcommand{\be}{\begin{equation}}
\newcommand{\eeq}{\end{equation}}
\newcommand{\bea}{\begin{eqnarray}}
\newcommand{\eea}{\end{eqnarray}}
\newcommand{\ba}{\begin{array}}
\newcommand{\ea}{\end{array}}

\newcommand{\ee}{\end{equation} }

\newcommand{\Tr}{\mathrm{Tr}\,}

\newcommand{\one}{{\rm 1\kern -.9mm l}}

\title{Shaping Lattice through irrelevant perturbation: Ising model}

\author{Armen Poghosyan}

\affiliation{Yerevan Physics Institute,\\
Alikhanian Br. 2, AM-0036 Yerevan, Armenia}
\emailAdd{armenpoghos@yerphi.am}

\abstract{The leading irrelevant perturbation, which controls the 
	deviation of critical square lattice Ising model with periodic boundary 
	conditions from its continuous CFT analog is identified. An explicit expression 
	for the coupling constant in terms of the anisotropy parameter is found. 
	We calculate the next to leading $\sim 1/N^2$ corrections to the spectrum
	on both lattice theory and the perturbed CFT sides for several classes 
	of states, always getting exact agreement. We discuss also how the 
    perturbing operators and the higher integrals of motion are related.}


\preprint{YerPhI/2019/08}
\begin{document}

\maketitle
\flushbottom
%\newpage
%
%
%
%
\section{Introduction}
It is well known that large scale behavior of many statistical systems 
near their critical points can be described by (Euclidean) Quantum Field 
Theories (QFT) (see e.g. \cite{domb1972phaseV6}). In such description 
many microscopic details of initial theory 
are washed out, so that various statistical systems may lead to the same 
continuous theory. In this respect it is interesting to investigate intermediate 
scales, where some subleading corrections (besides leading finite size 
effects) to the QFT description still are 
noticeable. Then it would be possible to clarify how specific microscopic 
structure of the system is reflected in this corrections. From QFT point of view 
such corrections can be described as perturbations 
by irrelevant operators which are allowed to carry nonzero spins, since the 
rotational invariance at small distances is violated 
\cite{domb1987phaseV11,CARDY1986200}. 
In this respect 
the two dimensional Ising model \cite{Ising:1925em} is an ideal object to investigate, since 
it is exactly integrable \cite{Onsager}. Moreover, for some boundary conditions 
all the eigenvalues of its transfer matrix are known exactly even for finite lattices (see 
e.g. \cite{baxter2007exactly} for toroidal boundary conditions, \cite{Brien} for 
several other cases).                         

The paper is organized as follows.

In section \ref{section2} we systematically investigate the eigenvalues 
of periodic critical Ising transfer-matrix in large $L$ limit ($L$ is 
the number of spins in a horizontal row). We show how $1/L$ terms 
exactly match with CFT prediction and compute the next $1/L^3$  corrections 
for certain families of eigenvalues. The $1/L^3$ corrections under discussion 
describe breakdown of rotational invariance due to lattice artifacts. 

In section \ref{section3} we address the question, how to perturb Ising 
CFT in order to get precisely those corrections which we obtained investigating 
lattice model. We were able to identify the perturbing 
fields and the respective coupling constants. Namely we show that the perturbing 
fields are the spin $4$ current and its antiholomorphic counterpart, which 
were introduced in the context of integrable structure of CFT long ago. Having 
non-zero spin, these fields break the rotational invariance, while being 
irrelevant, they slightly correct the large distance behavior in a way, to 
mimic the lattice result.

Finally we end up with a summary of our results and discuss the possibility  
of generalization for full spectrum and for higher order corrections.    
\section{Eigenvalues of the transfer matrix}\label{section2} 
\begin{center}
	\begin{figure}
		\center
		\includegraphics[bb =4cm 1.5cm 30cm 18cm,clip=true, width=1
		\textwidth]{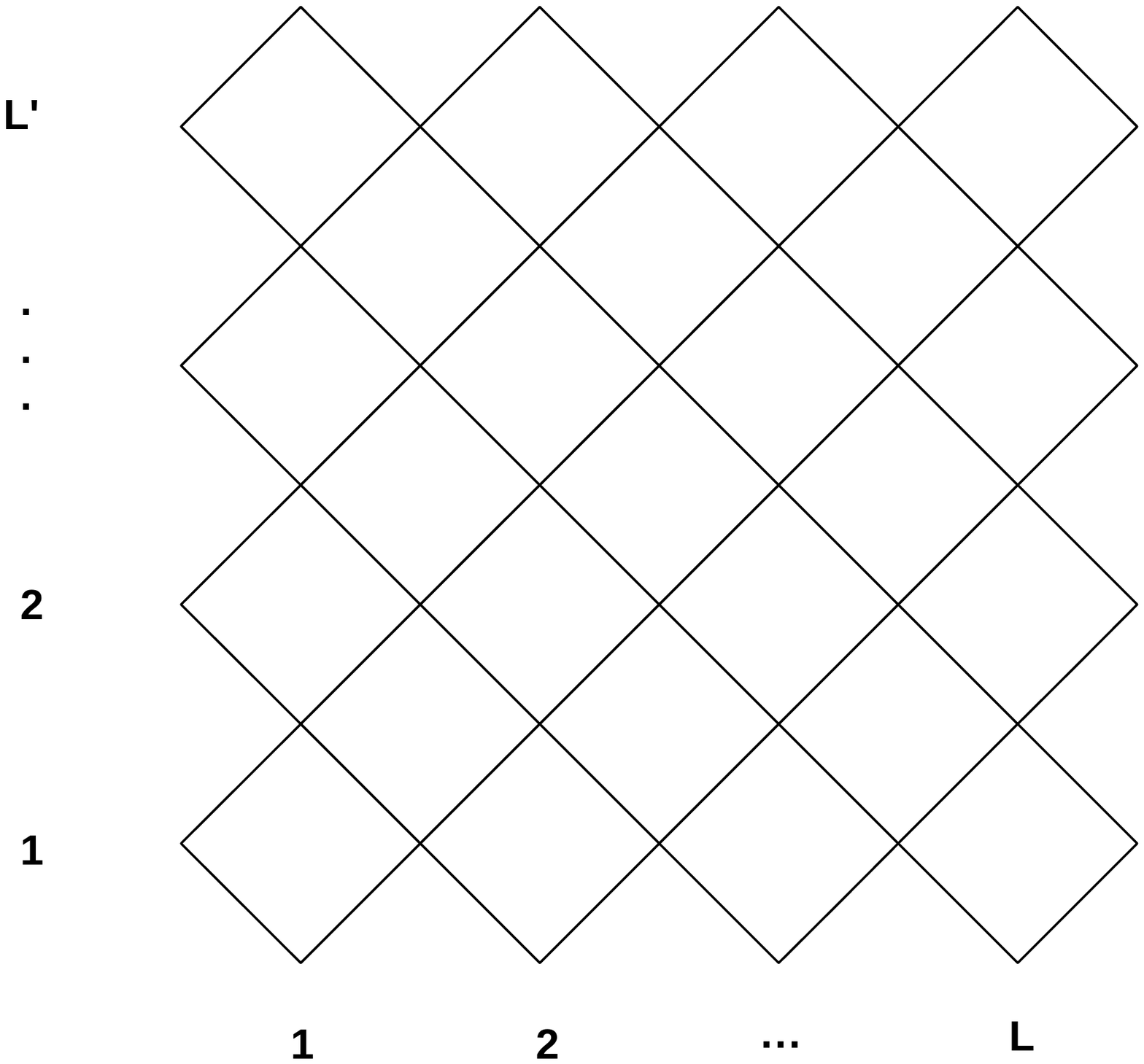}
		\caption{Periodic Ising Lattice. The leftmost (lowermost) 
		spins are identified with corresponding rightmost 
		(uppermost) spins.}
		\label{lattice}
	\end{figure}
\end{center}

Consider square lattice Ising model. We'll adopt the "$45$ degree rotated" version 
presented in great details e.g. in Baxter's seminal book \cite{baxter2007exactly}. 
The lattice consists of $L$  vertical columns of $L^\prime$ faces, or equivalently,
of $L^\prime$ horizontal rows of $L$ faces (see Fig.\ref{lattice}). We will consider 
periodic boundary condition in both, horizontal and vertical directions, so 
that the $L+1$-th column is identified with the first column and the  $L^\prime +1$-th 
row with the first one.
The partition function of the theory is given by
\bea
Z=\sum_{\{\mathbf{\sigma}\}}\exp \left(J\sum_{\langle i,j\rangle}
\sigma_i\sigma_j+K\sum_{\langle k,l\rangle}
\sigma_i\sigma_j
\right),
\eea 
where $\langle i,j\rangle$ belong to the set of SW-NE, and
$\langle k,l\rangle$ to the NW-SE edges. In what follows, we will 
restrict ourselves to the case of critical Ising model. The well 
known criticality condition $\sinh 2J \sinh 2K=1$ can be conveniently 
parameterized by a single parameter $u$ by setting
\bea
\sinh 2J=\cot 2u \,;\qquad  \sinh 2K=\tan 2u\,,
\label{JK_weights}
\eea 
with $0<u<\pi/4$. Sometimes the parameter $u$ is referred as the anisotropy.
The value $u=\pi/8$ corresponds to the isotropic case.    

\begin{center}
	\begin{figure}
		\center
		\includegraphics[bb = 0cm 7.5cm 28cm 15cm,clip=true, width=1
		\textwidth]{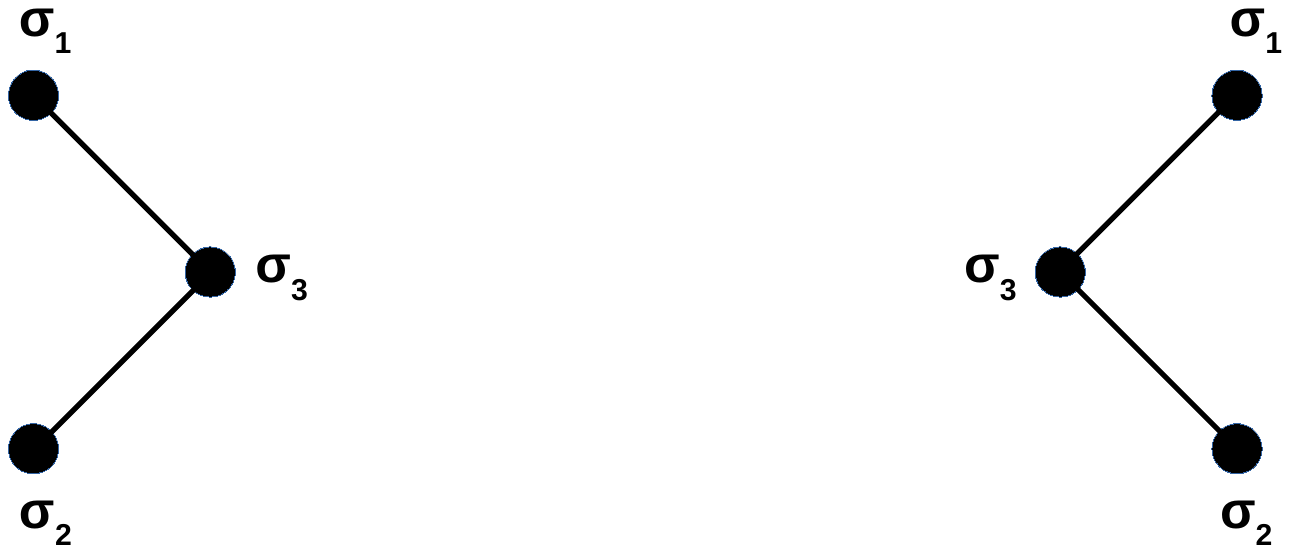}
		\caption{Two basic configurations of three Ising spins.}
		\label{w_weights}
	\end{figure}
\end{center}

To define the transfer matrix of the model, let us denote 
the Boltzmann weights of two basic types of three spin configurations shown in 
Fig.\ref{w_weights} as
\bea
W\left(
\begin{array}{ll}
	\sigma_1&\\[-10pt]
	& \sigma_3\\[-10pt]
	\sigma_2&
\end{array}
\right)
\qquad \text{and}\qquad
W\left(
\begin{array}{ll}
	&\sigma_1\\[-10pt]
	\sigma_3&\\[-10pt]
	&\sigma_2
\end{array}
\right)\,,\nonumber
\eea
where $\sigma_i=\pm1$ are the Ising spins. Thus, the arrangement 
of spins in the argument of $W$ follows to the geometric pattern of their 
locations on the lattice. 
These Boltzmann weights are explicitly  given by
\bea
W\left(
\begin{array}{ll}
	\sigma&\\[-10pt]
	& \sigma\\[-10pt]
	\sigma&
\end{array}
\right)&=&W\left(
\begin{array}{ll}
	&\sigma\\[-10pt]
	\sigma&\\[-10pt]
	&\sigma
\end{array}
\right)=\sqrt{2} \cos (u) \cos \left(\frac{\pi }{4}-u\right)\,,
\nonumber\\
W\left(
\begin{array}{ll}
	\sigma&\\[-10pt]
	& -\sigma\\[-10pt]
	\sigma&
\end{array}
\right)&=&
W\left(
\begin{array}{lr}
	&\sigma\\[-10pt]
	-\sigma&\\[-10pt]
	&\sigma
\end{array}
\right)=\sqrt{2} \sin (u) \sin \left(\frac{\pi }{4}-u\right)\,,\nonumber\\
W\left(
\begin{array}{rl}
	\sigma&\\[-10pt]
	& \sigma\\[-10pt]
	-\sigma&
\end{array}
\right)&=&W\left(
\begin{array}{cr}
	&-\sigma\\[-10pt]
	\sigma&\\[-10pt]
	&\sigma
\end{array}
\right)
=\sqrt{2} \sin (u) \cos \left(\frac{\pi }{4}-u\right)\nonumber\\
W\left(
\begin{array}{rc}
	-\sigma&\\[-10pt]
	& \sigma\\[-10pt]
	\sigma&
\end{array}
\right)&=&W\left(
\begin{array}{rr}
	&\sigma\\[-10pt]
	\sigma&\\[-10pt]
	&-\sigma
\end{array}
\right)
=\sqrt{2} \sin \left(\frac{\pi }{4}-u\right) \cos (u)\,,
\label{w_weights_formulae}
\eea 
where $\sigma=\pm 1$ and $u$ is the anisotropy parameter.
Note that to get (\ref{w_weights_formulae}) from (\ref{JK_weights}), 
one should include an overall extra factor $\frac{1}{2} \sqrt{\sin 4 u}$. 
Such shift in vacuum energy is convenient particularly because the 
transfer matrix, defined below, becomes the one-step shift operator 
at the values $u=0$ and $u=\frac{\pi}{4}$.  

\begin{center}
	\begin{figure}
		\center
		\includegraphics[bb = 0cm 7.5cm 28cm 13.5cm,clip=true, width=1\textwidth]{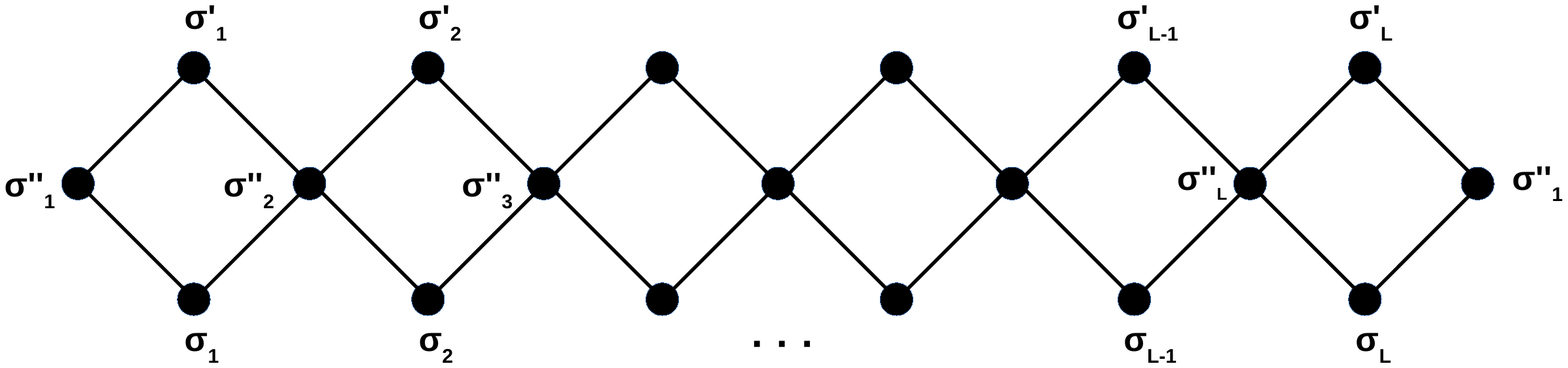}
		\caption{The transfer matrix. Sum over all middle spins $\sigma''$ is assumed }
		\label{tm}
	\end{figure}
\end{center}

After imposing periodic boundary condition in horizontal direction for 
the transfer-matrix we get (See Fig.\ref{tm})
\bea
T_{\mathbf{\sigma},\mathbf{\sigma}'}=\sum_{\mathbf{\sigma}''} 
W\left(
\begin{array}{ll}
	&\sigma_1'\\[-10pt]
	\sigma_1''&\\[-10pt]
	&\sigma_1
\end{array}
\right)
W\left(
\begin{array}{ll}
	\sigma_1'&\\[-10pt]
	& \sigma_2''\\[-10pt]
	\sigma_1&
\end{array}
\right)
W\left(
\begin{array}{ll}
	&\sigma_2'\\[-10pt]
	\sigma_2''&\\[-10pt]
	&\sigma_2
\end{array}
\right)\cdots 
W\left(
\begin{array}{ll}
	\sigma_L'&\\[-10pt]
	& \sigma_1''\\[-10pt]
	\sigma_L&
\end{array}
\right)
\label{transfer-matrix}
\eea 
The exact eigenvalues of this transfer-matrix can be found in 
\cite{baxter2007exactly}.
We will use a slightly modified version of the explicit expressions, 
presented in \cite{Brien} based on more general treatment of \cite{Behrend:1995zj}
 (see also later works \cite{Behrend1,Pearce2001,Chui}).  
There are two sets of eigenvalues, 
distinguished by the parameter $r=\pm 1$ and denoted by $\Lambda^{(\pm)}$ respectively. 
To be precise, let me 
warn the reader, that the quantities $\Lambda^{(\pm)}$ introduced below, 
actually are the {\it square roots} of transfer matrix eigenvalues. 
Nevertheless with a slight abuse of terminology, they will be referred simply 
as eigenvalues. 

For $r=1$ the set of eigenvalues are given explicitly as \cite{Brien}
($\lfloor x\rfloor$ stands for the largest integer not 
exceeding $x$):
\bea
\Lambda^{(+)}(\mu_k,\bar{\mu}_k)=
\sqrt{2} \left(2 e^{2 i u+\frac{i \pi }{4}}\right)^{-L} \prod _{k=1}^{\lfloor 
\frac{L}{2}\rfloor} \left(i \mu_k  \tan \left(\frac{\pi  (2 k-1)}{4 L}\right)
+e^{4 i u}\right)\nonumber\\
\prod _{k=1}^{\lfloor \frac{L+1}{2}\rfloor} \left(i \cot \left(\frac{\pi  
(2 k-1)}{4 L}\right)+e^{4 i u} \bar{\mu}_k \right),
 \label{+eigen}
\eea
where $\mu_k,\,\, \bar{\mu}_k\in \{+1,-1\}$ are subject to the constraint 
\bea
\prod_{k=1}^{\lfloor \frac{L}{2}\rfloor}\mu_k \prod_{k=1}^{\lfloor 
\frac{L+1}{2}\rfloor}\bar{\mu}_k=1 
\label{constraint}
\eea
For $r=-1$ we have another set of eigenvalues \cite{Brien}:
\bea
\Lambda^{(-)}(\mu_k,\bar{\mu}_k)=
\sqrt{1+L} \left(2 e^{2 i u+\frac{i \pi }{4}}\right)^{-L} \prod _{k=1}^{\lfloor 
	\frac{L}{2}\rfloor} \left(i \mu_k  \tan \left(\frac{\pi  k}{2 L}\right)+e^{4 i u}\right)\nonumber\\
\prod _{k=1}^{\lfloor \frac{L-1}{2}\rfloor} \left(i \cot \left(\frac{\pi  k}
{2 L}\right)+e^{4 i u} \bar{\mu}_k \right);
\label{-eigen}
\eea
where again  $\mu_k,\,\, \bar{\mu}_k\in \{+1,-1\}$, but there is no constraint anymore.

Notice that altogether we get $2^{L-1}+2^{L-1}=2^L$ eigenvalues, which exactly matches the 
size of transfermatrix (\ref{transfer-matrix}).  
\subsection{Leading Eigenvalues}
The leading eigenvalues in both sectors correspond to the case when all 
 $\mu_k=+1$, $\bar{\mu}_k=+1$. The expressions for these eigenvalues can be simplified 
 and represented as
\bea
 \label{simp_eigen_1}
 \left(\Lambda^{(+)}_{max}\right)^2=\prod _{k=0}^{L-1} \left(\sin (4 u) \sin \left(\frac{\pi  k}{L}
 +\frac{\pi }{2 L}\right)+1\right)
 \eea 
 and
\bea 
 \label{simp_eigen_2}
 \left(\Lambda^{(-)}_{max}\right)^2=\prod _{k=0}^{L-1} \left(\sin (4 u) \sin 
 \left(\frac{\pi  k}{L}\right)+1\right)
\eea 
\subsection{CFT prediction}
From now on instead of parameters $L$ and $L^\prime$ we'll use the even integers 
\[
N=2 L\quad \text{and}\quad M=2 L^\prime,
\]
which are the numbers of vertical 
and horizontal "zigzagging" columns and rows of the lattice respectively (see Fig.\ref{lattice}).
Conformal field theory predicts that the logarithm of transfer matrix eigenvalues in large 
$N$, $M$ limit behave as \cite{PhysRevLett.56.742, PhysRevLett.56.746}
\bea 
M \log \Lambda \sim -\frac{N M}{2}\, f_{bulk}+\left(\Delta-\frac{c}{24}\right)\log q+
\left(\bar{\Delta}-\frac{c}{24}\right)\log \bar{q}  ,
\label{CFT_prediction}
\eea
where $f_{bulk}$ is the free energy per site (the number of sights in our lattice 
is $\frac{NM}{2}$), 
$\Delta$, $\bar{\Delta}$ are the left and right dimensions of the conformal field 
which creates the corresponding eigenstate from the vacuum. The  parameter $q=\exp 2\pi i \tau $, 
where $\tau $ is modulus 
of the torus on which our 2d CFT lives, should be identified as 
\bea
2 \pi i \tau= \log q=-\frac{2 \pi i M\,e^{-4 i u}}{N}
\label{q} 
\eea 
Finally, $c$ is the Virasoro central charge, which for our case of Ising 
model, assumes the value $c=1/2$.  

Notice also that besides finite size scaling, CFT predicts also the 
universal amplitude ratios (see e.g. \cite{Salas:1999qh} and references therein).
\subsection{Large $N$ expansion of leading eigenvalues}
Denote
\bea 
f(x)=\log (1+\sin (4u) \sin (x))
\eea 
and $h=\frac{2\pi}{N}$. Due to Euler-Maclaurin summation formula ($B_n(\alpha) $
are the Bernoulli polynomials)  from eq. (\ref{simp_eigen_1}) we get
\bea 
&&\log\Lambda^{(+)}_{max}+\frac{N}{2}\, f_{bulk}=\sum_{n=1}^\infty  \frac{B_n\left(\frac{1}{2}\right) \left(f^{(n-1)}(\pi)
	-f^{(n-1)}(0)\right) h^{n-1}}{n!}\nonumber\\
&&=\frac{\pi  \sin (4 u)}{12 N}+\frac{7 \pi ^3 \left(2 \sin (4 u)-4 \sin ^3(4 u)\right)}{1440 N^3}+
O(1/N^5)
\label{log_leading}
\eea 
and from (\ref{simp_eigen_2}) 
\bea 
&&\log\Lambda^{(-)}_{max}+\frac{N}{2} f_{bulk}=\sum_{n=1}^\infty  \frac{B_n\left(0\right) \left(f^{(n-1)}(\pi)
	-f^{(n-1)}(0)\right) h^{n-1}}{n!}\nonumber\\
&&=-\frac{\pi  \sin (4 u)}{6 N}-\frac{\pi ^3 \left(2 \sin (4 u)-4 \sin ^3(4 u)\right)}{180 N^3}+
O(1/N^5)
\label{log_Nleading}
\eea 
The bulk free energy is given by
\bea
f_{bulk}=-\int_0^\pi f(x)\,\frac{dx}{2\pi}.
\eea
Taking into account that the leading eigenvalue should correspond to the identity operator with 
conformal dimensions $\Delta=\bar{\Delta}=0$, consistency of the leading terms 
in (\ref{log_leading}), (\ref{log_Nleading}) with CFT prediction (\ref{CFT_prediction}) 
immediately fixes the value of central charge 
$c=1/2$ and identifies the torus parameter $q$ with the expression 
(\ref{q}). Besides for the dimensions of the conformal field corresponding to 
the eigenvalue $\Lambda^{(-)}_{max}$ we get $\Delta=\bar{\Delta}=1/16$, which are the correct
dimensions of Ising spin field.
\subsection{Partition function on a large torus and Ising fermions}
\label{2.4}
For given $p$, denote $\Lambda^{(+)}_p$  ($\bar{\Lambda}^{(+)}_p$) the rhs 
of (\ref{+eigen}) for the case when $r=1$ with all $\mu$'s and $\bar{\mu}$'s set to $1$ with a single 
exception, namely $\mu_p=-1$  ($\bar{\mu}_p=-1$). Analogously, using  (\ref{-eigen}), we define 
$\Lambda^{(-)}_p$, $\bar{\Lambda}^{(-)}_p$ for the case when $r=-1$. 
In large $N$ limit the impact of flipping the sign at position $p$ are 
given by ratios 
\bea
\log \frac{\Lambda^{(+)}_p}{\Lambda^{(+)}_{max}}&=&-\frac{i \pi  (2 p-1) e^{-4 i u}}{N}
-\frac{i \pi ^3 (2 p-1)^3 e^{-4 i u} \left(1-e^{-8 i u}\right)}{12 N^3}+O\left(1\over N^5\right)
\nonumber\\
\log \frac{\bar{\Lambda}^{(+)}_p}{\Lambda^{(+)}_{max}}&=&
\frac{i \pi  (2 p-1) e^{4 i u}}{N}+
\frac{i \pi ^3 (2p-1)^3 e^{4 i u} \left(1-e^{8 i u}\right)}{12 N^3}+O\left(1\over N^5\right)
\label{sector_plus_expansion}
\eea 
and
\bea
\log \frac{\Lambda^{(-)}_p}{\Lambda^{(-)}_{max}}&=&-\frac{2 i \pi  p e^{-4 i u}}{N}
-\frac{2 i \pi ^3 p^3 e^{-4 i u} \left(1-e^{-8 i u}\right)}{3 N^3}+O\left(1\over N^5\right)
\nonumber\\
\log \frac{\bar{\Lambda}^{(-)}_p}{\Lambda^{(-)}_{max}}
&=&\frac{2i \pi  p e^{4 i u}}{N}+\frac{2i \pi ^3 p^3 e^{4 i u} 
\left(1-e^{8 i u}\right)}{3 N^3}+O\left(1\over N^5\right)
\label{sector_minus_expansion}
\eea 
It is important to notice that due to factorized structure of eigenvalues (\ref{+eigen}) 
and (\ref{-eigen}), if there are {\it several} indices $p_1,p_2,\cdots $,
such that the respective $\mu$'s or $\bar{\mu}$'s assume the value $-1$, the logarithm of ratios 
of eigenvalues will be given by the same expressions (\ref{sector_plus_expansion}) 
and  (\ref{sector_minus_expansion})  summed over above specified indices p.  

Consider thermodynamic 
limit when $M$, $N$ are sent to infinity while keeping their ratio fixed. Notice that $q$, $\bar{q}$ 
remain finite as seen from (\ref{q}). Keeping the leading terms only, we 
can conveniently  rewrite expressions (\ref{sector_plus_expansion}), (\ref{sector_minus_expansion})  
in view of eq. (\ref{q}) as
\bea
\left(\frac{\Lambda^{(+)}_p}{\Lambda^{(+)}_{max}}\right)^M \sim q^{p-\frac{1}{2}}; \,\,\, 
\left(\frac{\Lambda^{(-)}_p}{\Lambda^{(+)}_{max}}\right)^M
\sim q^{\bar{p}-\frac{1}{2}};\,\,\,
\left(\frac{\bar{\Lambda}^{(+)}_p}{\Lambda^{(-)}_{max}}\right)^M \sim q^{p};\,\,\, 
\left(\frac{\Lambda^{(-)}_p}{\Lambda^{(-)}_{max}}\right)^M
\sim \bar{q}^{p}\qquad
\eea
In other words, the sign flip of $\mu_p$ ( $\bar{\mu}_p$) 
in partition sum costs a multiplier $q^{p-\frac{1}{2}}$ ($\bar{q}^{p-\frac{1}{2}}$) 
in the $r=1$ sector and a multiplier $q^{p}$ ($\bar{q}^{p}$) in the sector $r=-1$.
We'll need also the contributions of the eigenvalues $\Lambda^{(\pm)}_{max}$ in partition
function. Again, dropping $O(1/N^3)$ terms, from eqs. (\ref{log_leading}), 
(\ref{log_Nleading}) we easily get
\[
\left(\Lambda^{(+)}_{max}\right)^M\sim e^{-MN f_{bulk}/2}
(q\bar{q})^{-\frac{1}{48}}; \qquad \left(\Lambda^{(-)}_{max}\right)^M\sim e^{-MN f_{bulk}/2}
(q\bar{q})^{-\frac{1}{48}+\frac{1}{16}}.
\] 
As a result, a generic eigenvalue, say in sector $r=-1$, specified by conditions $\mu_{p_1}=\mu_{p_2}
=\cdots =\mu_{p_R}=-1$,  $\bar{\mu}_{\tilde{p}_1}=\bar{\mu}_{\tilde{p}_2}
=\cdots =\bar{\mu}_{\tilde{p}_L}=-1$\footnote{$L$, the number of parameters  $\bar{\mu}$ taking 
the value $-1$ should not be confused with the number of rows of faces introduced earlier. 
In fact the latter does not appear in the paper any more.} 
with all other $\mu$'s and $\bar{\mu}$'s taking the 
value $1$, will contribute a term 
\[
e^{-MN f_{bulk}/2}
(q\bar{q})^{-\frac{1}{48}+\frac{1}{16}} q^{p_1}q^{p_2}\cdots q^{p_R} 
\bar{q}^{\tilde{p}_1}\bar{q}^{\tilde{p}_2}\cdots \bar{q}^{\tilde{p}_L}
\] 
Similarly, the contribution of a generic eigenvalue from the sector $r=-1$ 
in partition function reads
\[
e^{-MN f_{bulk}/2}
(q\bar{q})^{-\frac{1}{48}+\frac{1}{16}} q^{p_1-1/2}q^{p_2-1/2}\cdots q^{p_R-1/2} 
\bar{q}^{\tilde{p}_1-1/2}\bar{q}^{\tilde{p}_2-1/2}\cdots \bar{q}^{\tilde{p}_L-1/2}.
\] 
Here we should keep in mind that due to the constraint (\ref{constraint}), 
the total number of minus signs, i.e. $R+L$ must be even. 
Now, by elementary considerations one can get convinced that the partition function 
on the torus in continuum limit can be represented as
\bea
Z\sim e^{-MN f_{bulk}/2}
(q\bar{q})^{-\frac{1}{48}}\hspace{8cm}\nonumber\\
\times\left( \frac{1}{2}\prod_{k=1}^\infty \left(1+q^{k-\frac{1}{2}}\right)
\left(1+\bar{q}^{k-\frac{1}{2}}\right)+ \frac{1}{2}\prod_{k=1}^\infty 
\left(1-q^{k-\frac{1}{2}}\right)
\left(1-\bar{q}^{k-\frac{1}{2}}\right)\right.
\nonumber\\
\left. +(q\bar{q})^{\frac{1}{16}} \prod_{k=1}^\infty \left(1+q^k\right)
\left(1+\bar{q}^k\right)\right).
\label{Z_cont}
\eea 
Note that the sum with factors $1/2$ on second line implements the constraint 
(\ref{constraint}), since the terms with a product of odd number of minus signs get canceled. 

Obviously, the form of (\ref{Z_cont}) reflects the well known fact that 
the Ising model in continuous limit is the theory of free fermions. The 
fermionic fields are single valued in vacuum sector (manifested by the 
half integer modes on second line of (\ref{Z_cont})),  while they are 
non-local with respect to the spin field   
and have integer modes as seen on third line of  (\ref{Z_cont}). 
Let us briefly recall few relevant facts about free fermion theory. We will simply 
state the results omitting details and proofs. The aim is to specify notations which will be used later on.   A comprehensive review of 2d fermion CFT can 
be found e.g. in book \cite{DiFrancesco:639405}.
The left and right moving fermion fields can be defined through mode expansions
\bea 
\psi(z)=\sum_{\nu} \frac{\psi_\nu}{z^{\nu+\frac{1}{2}}}\\
\bar{\psi}(\bar{z})=\sum_{\nu} \frac{\bar{\psi}_\nu}{\bar{z}^{\nu+\frac{1}{2}}}
\eea
where sum is over all half integers or over all integers, depending whether 
it acts in vacuum sector (usually referred as Neveu-Schwartz sector) or on spin
(Ramond) sector. The expansion modes satisfy the anticommutation relations
\bea 
\{\psi_\nu ,\psi_\mu\} =\delta_{\nu+\mu ,0}\,;\qquad 
\{\bar{\psi}_\nu ,\bar{\psi}_\mu\} =\delta_{\nu+\mu ,0}
\eea 
In terms of fermion modes the sector $r=1$ is spanned on states
\bea
\prod_{k=1}^\infty \left(\psi_{-k+\frac{1}{2}}\right)^{\varepsilon_{k-\frac{1}{2}}}
\prod_{k=1}^\infty \left(\bar{\psi}_{-k+\frac{1}{2}}\right)^{\bar{\varepsilon}_{k-\frac{1}{2}}}
|0\,;0\rangle
\label{states_p}
\eea
where $|0\,;0 \rangle $ is the vacuum state with dimensions $\Delta=\bar{\Delta}=0$ 
and the occupation numbers $\varepsilon_\nu$, $\bar{\varepsilon}_\nu$ assume values $0$ or $1$. 
The relation of (\ref{states_p}) to the eigenstates of Ising transfer matrix is very simple: 
if an occupation number $\varepsilon_{k-1/2}=1$ ($\varepsilon_{k-1/2}=0$) then $\mu_k=-1$ ($\mu_k=1$). 
The analogous relation holds also for quantities  $\bar{\varepsilon}$ and $\bar{\mu}$.
As a consequence, the constraint (\ref{constraint}) requires, that the total fermion number 
\bea
N_f =\sum_{\nu} \left(\varepsilon_{\nu}+\bar{\varepsilon}_{\nu}\right)\equiv 
N_L+N_R
\label{f_number}
\eea 
must be even. Evidently there are two alternatives here. Either both $N_L$ 
and $N_R$ are even or both are odd. It is easy to see that the former case 
corresponds to the Virasoro module created from the vacuum  $|0\,;0\rangle $ 
while the latter one to the module created from primary state (related to the 
energy density) $|\frac{1}{2}\,;\frac{1}{2} \rangle $.

Similarly the Ramond subspace is spanned over the states
\bea
\prod_{k=1}^\infty \left(\psi_{-k}\right)^{\varepsilon_{k}}
\prod_{k=1}^\infty \left(\bar{\psi}_{-k}\right)^{\bar{\varepsilon}_{k}}
\left|\frac{1}{16}\,; \frac{1}{16}\right\rangle\,
\label{states_a}
\eea 
which exactly match the states of the sector $r=-1$ in lattice side. 
It is not difficult to calculate the torus partition function of this free 
fermion theory
\[
Z_{ferm}=\Tr q^{L_0-\frac{c}{24}}\bar{q}^{\bar{L}_0-\frac{c}{24}}\,,
\] 
where $L_0$, $\bar{L}_0$ are left and right Virasoro 0-modes, $q$ and $\bar q$,
as earlier, are the torus parameter and its conjugate and $c=1/2$ is the Virasoro central charge.
The trace is over the states (\ref{states_a}) and (\ref{states_p}) subject to the constraint 
$N_f=0\mod 2$ (see eq. (\ref{f_number})). The result  besides the non universal factor 
$\exp (-NM f_{bulk})$ exactly matches the the partition sum (\ref{Z_cont}).
\section{Irrelevant perturbation of Ising CFT}\label{section3}
Let us introduce coordinate $\zeta =x+i y$ on cylinder ($\zeta\sim \zeta+2\pi $), which is related to 
the coordinate $z$ on plane through exponential map $z=\exp \zeta $. The least irrelevant perturbation 
of Ising CFT from the conformal family of identity operator, responsible for deviation from conformal theory, is 
(see \cite{Cardy:1986ie}, \cite{CARDY1986200}, \cite{Zamolodchikov:1987ti}, \cite{domb1987phaseV11}) 
\bea 
H_{int}=\int \left(g T_{cyl}^2(\zeta)+\bar{g}\bar{T}_{cyl}^2(\bar{\zeta})\right)\frac{dy}{2\pi}
\eea 
where $T_{cyl}^2$ and its antiholomorphic counter part are regularized squares of the energy 
momentum tensor. More precisely
\bea 
T_{cyl}^2(\zeta)\equiv \oint\frac{ T_{cyl}(\zeta ')T_{cyl}(\zeta )d\zeta '}{2\pi i(\zeta '-\zeta)}
\eea
where integration over $\zeta '$ is along a small  contour surrounding $\zeta$ anticlockwise. 
Using transformation rule of stress-energy tensor from plain to cylinder
\bea 
T_{cyl}(\zeta)=z^2 T(z)-\frac{c}{24} ,
\eea 
where the second term comes from Schwarzian derivative, one can see that in terms of 
conventional Virasoro modes the interaction Hamiltonian becomes
\bea 
H_{int}=g\left(2\sum_{n=1}^{\infty}L_{-n}L_n+L_0^2-\frac{c+2}{12}\,L_0+\frac{c(22+5c)}{2880}\right)\nonumber\\
+\bar{g}\left(2\sum_{n=1}^{\infty}\bar{L}_{-n}\bar{L}_n+\bar{L}_0^2-\frac{c+2}{12}\,\bar{L}_0+\frac{c(22+5c)}{2880}\right)\,.
\label{H_int}
\eea 
Note that (\ref{H_int}) is a combination of
higher integrals of motion $I_3$ and $\bar{I}_3$ \cite{Sasaki:1987mm,Eguchi:1989hs,Bazhanov:1994ft}
\[
H_{int}=gI_3+\bar{g}\bar{I}_3
\] 
Remined also that the 
unperturbed Hamiltonian coincides with 
\[
H_0=L_0+\bar{L}_0-\frac{c}{12}\equiv I_1+\bar{I}_1
\] 
Since the 
commutator $[I_1,I_3]=0$, the perturbation theory appears to be remarkably simple. 
In fact, the calculations presented in the remaining part of the paper, strongly suggest that 
eigenstates of the transfer-matrix (\ref{transfer-matrix}) simultaneously diagonalize all 
operators $I_1$, $I_3$, $\bar{I}_1$ and  $\bar{I}_3$. It is expected that this 
feature is generic, and higher order calculations would involve further integrals of 
motion $I_5$, $I_7$, ... as well, thus making contact between integrable structures 
of the lattice model and CFT (see \cite{Morin_Duchesne_2016} for discussion of the dimer model 
from this perspective).

Representation of the interaction Hamiltonian in form (\ref{H_int}) is very convenient 
for calculation of various matrix elements. In particular it is 
straightforward (for arbitrary $c$) to get:
\bea 
\frac{\langle \Delta\,;\bar{\Delta}|L_1^p \bar{L}_1^{\tilde{p}} 
H_{int}L_{-1}^p\bar{L}_{-1}^{\tilde{p}}|\Delta\,;\bar{\Delta}\rangle }
{\langle \Delta\,;\bar{\Delta}|L_1^p\bar{L}_{-1}^{\tilde{p}} L_{-1}^p
\bar{L}_{-1}^{\tilde{p}}| \Delta\,;
\bar{\Delta}\rangle }=g M(\Delta,p)+\bar{g} M(\bar{\Delta},\tilde{p})
\label{matrix_elm_L}
\eea 
where
\bea 
M(\Delta,p)= \left(\frac{c}{24}\right)^2+\frac{11 c}{1440}
+(\Delta +p) \left(-\frac{c}{12}+\Delta 
+\frac{p (2 \Delta +p)(5 \Delta +1)}{(\Delta +1)
(2 \Delta +1)}-\frac{1}{6}\right)\quad 
\label{M_delta_p}
\eea
This result was found by a different technique long ago in 
\cite{reinicke87}.  
Another set of matrix elements for Ising model will be \cite{pki,PhysRevE.96.062127}
(now the central charge is specified to $c=1/2$)
\bea 
\frac{\langle \frac{1}{16}\,;\frac{1}{16}|\psi_{p} H_{int}\psi_{-p}|
\frac{1}{16}\,;\frac{1}{16}\rangle }{\langle \frac{1}{16}\,;
\frac{1}{16}|\psi_{p} \psi_{-p}|
\frac{1}{16}\,;\frac{1}{16}\rangle }=g M(p)+\bar{g} M(0)
\label{matrix_elm_psi} 
\eea 
where
\bea 
M(p)= \frac{7 p^3}{6}-\frac{7}{1440}
\label{M_p}
\eea 
\subsection{Identification of coupling constants}
\label{3.1}
Comparing the leading term of (\ref{log_leading}) with (\ref{CFT_prediction}) 
and taking into account that for vacuum state $\Delta=\bar{\Delta}=0$, we easily 
verify that the values of parameters $q$, $\bar{q}$ are given by eq. (\ref{q}) 
and that central charge $c=\frac{1}{2}$. 
It is convenient to represent (\ref{log_leading}) (included the
second, next to leading term) as ($c.c.$ stands for complex conjugate)
\bea
\frac{\pi  \sin (4 u)}{12 N}&+&\frac{7 \pi ^3 \left(2 \sin (4 u)-
4 \sin ^3(4 u)\right)}{1440 N^3}\nonumber\\
&=&\left(-\frac{1}{48}+\frac{7 \pi ^2}{5760 N^2}\left(1-e^{-8iu}\right)\right)\frac{\log q}{M}+c.c.
\eea 
The term 
\bea 
\delta E_0=\frac{7 \pi ^2}{5760 N^2}\left(1-e^{-8iu}\right)
\eea 
 can be interpreted as a shift of
(holomorphic part of) the vacuum state energy due to perturbation by $H_{int}$. This leads to 
the relation
\bea 
g M(0,0)=\frac{7 \pi ^2}{5760 N^2}\left(1-e^{-8iu}\right)\,.
\eea 
Since, due to (\ref{M_delta_p})
\[
M(0,0)=\frac{c (5 c+22)}{2880}=\frac{49}{11520}
\] 
for the coupling constant we get
\bea 
g=\frac{2 \pi^2}{7 N^2}\left(1-e^{-8iu}\right)\,.
\label{g_value}
\eea
Analogously, the conjugate coupling
\bea 
\bar{g}=\frac{2 \pi^2}{7 N^2}\left(1-e^{8iu}\right)\,.
\label{gbar_value}
\eea 
Now let us consider in some details the isotropic case $u=\frac{\pi}{8}$. In 
this case we have  
\[
g=\bar{g}=\frac{4 \pi^2}{7 N^2}
\] 
It is subtle but possible to compare this result with the coupling constant
\[
g_l=-\frac{1}{28\pi}\,,
\] 
obtained in \cite{Izmailian2001} for the (not rotated) square lattice critical Ising model. 
To make a correct comparison one should take into account that in current paper we have chosen
a natural from CFT point of view energy  normalization such that the level spacing in leading 
order is equal to $1$, while in 
\cite{Izmailian2001} it is equal to $\frac{\tilde{N}}{2 \pi}$ (to avoid confusion we denote 
the parameters $N$, $M$, of  \cite{Izmailian2001} by 
$\tilde{N}$, $\tilde{M}$). Besides, $\tilde{N}$ should be rescaled by the geometric 
factor $\sqrt{2}$ (since the number of sites $\frac{NM}{2}$ should be identified with 
$\tilde{N}\tilde{M}$). Finally a factor 
$-1$ arises due to rotation of lattice by $\theta=\frac{\pi}{4} $. Indeed, 
the spin of the perturbing field is $s=4$, hence we get the phase factor
\[
e^{i \theta s}=-1\,.
\] 
Carefully picking up all the factors arising from above considerations one 
arrives at the consistency condition
\[
g=-\frac{(2 \pi)^3}{\tilde{N}^2}g_l
\] 
which, obviously is satisfied.

In what follows, besides the vacuum state,  we'll examine infinitely many other 
states and will demonstrate how the values (\ref{g_value}), (\ref{gbar_value}) of the
coupling constants reproduces correct energy shifts.  
\subsection{Energy corrections for the states $L_{-1}^{p-1}\bar{L}_{-1}^{\tilde{p}-1}|1/2;1/2\rangle $}
\label{3.2}
It is straightforward to deduce from (\ref{sector_plus_expansion}), (\ref{log_leading}) that the large $N$ 
expansion of the transfer-matrix eigenvalues in $r=1$ sector with two sign flips $\mu_p=-1$ and 
$\bar{\mu}_{\tilde{p}}=-1$ can be represented as  
\bea 
&&M \log\Lambda^{(+)}_{p,\tilde{p}}+\frac{N M}{2}\, f_{bulk}=\nonumber\\
&&=\left(-\frac{1}{48}+\frac{1}{2}+p-1+\frac{\pi ^2\left(1-e^{-8iu}\right)}{5760 N^2}\left(7
+1920\left(p-1/2\right)^3\right)\right)\log q \nonumber\\
&&+\left(-\frac{1}{48}+\frac{1}{2}+\tilde{p}-1+\frac{\pi ^2\left(1-e^{8iu}\right)}{5760 N^2}\left(7
+1920\left(\tilde{p}-1/2\right)^3\right)\right)\log \bar{q} \nonumber\\
&&+O(1/N^5)
\label{log_lambdap_pp}
\eea
The leading terms in above expressions are displayed in such specific way to emphasize the structure
\[
-\frac{c}{24} +\Delta+ l\,
\] 
where $\Delta $ is the dimension of respective {\it primary} field and $l$ is 
the excitation level.
According to eq. (\ref{states_p}), the states under consideration should be identified 
with $\psi_{-p+\frac{1}{2}}\bar{\psi}_{-\tilde{p}+\frac{1}{2}}|0\,;0 \rangle$.
Using commutation relation
\bea 
[ L_n,\psi_\nu ]=-\left(\frac{n}{2}+\nu\right)\psi_{n+\nu}
\eea 
it is easy to show that 
\bea 
L_{-1}^p\bar{L}_{-1}^{\tilde{p}}\left|1/2\,;1/2\right\rangle \sim
\psi_{-p+\frac{1}{2}}\bar{\psi}_{-\tilde{p}+\frac{1}{2}}|0\,;0 \rangle 
\eea 
Comparing (\ref{matrix_elm_L}) with subleading terms of 
(\ref{log_lambdap_pp} ) we see that consistency with perturbation 
theory requires the equality
\bea
g M(1/2,p)=\frac{\pi ^2\left(1-e^{-8iu}\right)}{5760 N^2}\left(7
+1920\left(p-1/2\right)^3\right)
\eea 
which is easily checked to be satisfied identically due to (\ref{g_value}) 
and (\ref{M_delta_p}).
\subsection{Energy corrections for the states $\psi_{-p}\bar{\psi}_{-\tilde{p}}|1/16\,;1/16 \rangle $} 
\label{3.3}
In this section we will examine the analogues infinite series of states in $r=-1$ sector. Again 
we will assume that $\mu_{p}=-1$, $\bar{\mu}_{\tilde{p}}=-1$ and all other $\mu\,,\bar{\mu}=1$.
In this case we get the large $N$ expansion
\bea 
&&M \log\Lambda^{(-)}_{p,\tilde{p}}+\frac{N M}{2}\, f_{bulk}=\nonumber\\
&&=\left(-\frac{1}{48}+\frac{1}{16}+p+\frac{\pi ^2\left(1-e^{-8iu}\right)}{720 N^2}\left(-1
+240p^3\right)\right)\log q \nonumber\\
&&+\left(-\frac{1}{48}+\frac{1}{16}+\tilde{p}+\frac{\pi ^2\left(1-e^{8iu}\right)}{720 N^2}\left(-1
+240 \tilde{p}^3\right)\right)\log \bar{q} \nonumber\\
&&+O(1/N^5)
\label{log_lambdam_pp}
\eea

In this case the perturbation theory requires (see  (\ref{matrix_elm_psi}) and 
\ref{sector_minus_expansion}):
\bea
g M(p)=\frac{\pi ^2\left(1-e^{-8iu}\right)}{720 N^2}\left(-1
+240p^3\right)
\eea 
which again is satisfied identically due to (\ref{g_value}) 
and (\ref{M_p}).
\subsection{An example of doubly degenerate states}
\label{3.4}
Now let us consider a case, when double degeneracy takes place. Namely 
we will consider two forth level  states
\bea 
|1\rangle \equiv \psi_{-7/2}\psi_{-1/2}|0;0\rangle\,; \qquad 
|2\rangle \equiv \psi_{-5/2}\psi_{-3/2}|0;0\rangle \,.
\eea
Though not obvious, a direct calculation,  
sketched below, shows that the matrix element of $H_{int}$ between 
these states is zero, and that the energy shifts for either of these states indeed 
agree with lattice prediction. 
 
To make actual calculations, let us remind that
\bea 
T(z)=-\frac{1}{2}:\psi(z)\partial \psi(z):
\eea 
where $::$ stands for normal ordering. In terms of modes this is equivalent to
\bea
L_k=\frac{1}{4}\sum_{\nu+\mu=k}(\mu-\nu):\psi_\nu\psi_\mu:
\eea 
As an example let us calculate the action of $L_{-2}^2$ on vacuum state:
\bea
&&L_{-2}^2|0;0\rangle=\nonumber\\
&&(\cdots \psi_{-\frac{7}{2}}\psi_{\frac{3}{2}}+\psi_{-\frac{5}{2}}\psi_{\frac{1}{2}}
+\psi_{-\frac{3}{2}}\psi_{-\frac{1}{2}})(\cdots+\psi_{-\frac{5}{2}}\psi_{\frac{1}{2}}
+\psi_{-\frac{3}{2}}\psi_{-\frac{1}{2}})|0;0\rangle =\nonumber\\
&&\frac{5}{4}\psi_{-\frac{7}{2}}\psi_{-\frac{1}{2}}|0;0\rangle
-\frac{3}{4}\psi_{-\frac{5}{2}}\psi_{-\frac{3}{2}}|0;0\rangle 
\equiv \frac{5}{4}|1\rangle -\frac{3}{4}|2\rangle \,.
\label{L-2-2}
\eea
Similarly we get
\bea
L_{-1}^2L_{-2}|0;0\rangle =3|1\rangle +|2\rangle\,.
\label{L-1-1-2}
\eea 
Inverting the relations (\ref{L-2-2}) and (\ref{L-1-1-2}) we obtain
\bea 
|1\rangle&=&\frac{2}{7}L_{-2}^2|0;0\rangle +\frac{3}{14}L_{-1}^2L_{-2}|0;0\rangle \,,\nonumber\\
|2\rangle&=&-\frac{6}{7}L_{-2}^2|0;0\rangle+\frac{5}{14}L_{-1}^2L_{-2}|0;0\rangle\,.
\eea
Thus calculation of matrix elements of the interaction Hamiltonian (\ref{H_int}) 
boils down to simple Virasoro algebra manipulations. Here are the results of 
calculations:
\bea 
\langle 1|H_{int}|1\rangle&=&\frac{577969}{11520}\, g+\frac{49}{11520} \,\bar{g}\nonumber\\
\langle 2|H_{int}|2\rangle&=&\frac{255409}{11520}\, g+\frac{49}{11520} \,\bar{g}\nonumber\\
\langle 1|H_{int}|2\rangle&=&\langle 2|H_{int}|1\rangle=0
\label{H_int_ij}
\eea
According to the rules established in subsection \ref{2.4} the state $|1\rangle $ corresponds 
to the transfer-matrix eigenvalue with sign flips $\mu_{1}=-1$, $\mu_{4}=-1$ 
and $|2\rangle $ corresponds to  $\mu_{2}=-1$, $\mu_{3}=-1$ (both states belong 
to the sector $r=1$). From (\ref{sector_plus_expansion}), (\ref{log_leading}) 
we get expansions 
\bea 
&&M \log\Lambda^{(+)}_{1,4}+\frac{N M}{2}\, f_{bulk}=\nonumber\\
&&=\left(-\frac{1}{48}+4+\frac{\pi ^2\left(1-e^{-8iu}\right)}
{N^2}\left(\frac{7}{5760}+\frac{1}{3}\left( 
\left(1-1/2\right)^3+\left(4-1/2\right)^3\right)\right)\right)\log q \nonumber\\
&&+\left(-\frac{1}{48}+\frac{7\pi ^2\left(1-e^{8iu}\right)}
{5760N^2}\right)\log \bar{q}+O(1/N^5)
\label{log_14}
\eea
\bea 
&&M \log\Lambda^{(+)}_{2,3}+\frac{N M}{2}\, f_{bulk}=\nonumber\\
&&=\left(-\frac{1}{48}+4+\frac{\pi ^2\left(1-e^{-8iu}\right)}
{N^2}\left(\frac{7}{5760}+\frac{1}{3}\left( 
\left(2-1/2\right)^3+\left(3-1/2\right)^3\right)\right)\right)\log q \nonumber\\
&&+\left(-\frac{1}{48}+\frac{7\pi ^2\left(1-e^{8iu}\right)}
{5760N^2}\right)\log \bar{q}+O(1/N^5)
\label{log_23}
\eea
The expressions (\ref{log_14}), (\ref{log_23}) are in perfect agreement 
with (\ref{H_int_ij}), since the relations
\bea 
\frac{577969}{11520}\, g&=&
\frac{\pi ^2\left(1-e^{-8iu}\right)}
{N^2}\left(\frac{7}{5760}+\frac{1}{3}\left( 
\left(1-1/2\right)^3+\left(4-1/2\right)^3\right)\right)=
\frac{82567}{5760}\frac{\pi ^2\left(1-e^{-8iu}\right)}
{N^2},
\nonumber\\
\frac{255409}{11520}\, g&=&
\frac{\pi ^2\left(1-e^{-8iu}\right)}
{N^2}\left(\frac{7}{5760}+\frac{1}{3}\left( 
\left(2-1/2\right)^3+\left(3-1/2\right)^3\right)\right)=
\frac{36487}{5760}\frac{\pi ^2\left(1-e^{-8iu}\right)}
{N^2},
\nonumber\\
\frac{49}{11520} \,\bar{g}&=&\frac{7\pi ^2\left(1-e^{8iu}\right)}
{5760N^2}
\eea 
in view of (\ref{g_value}), are satisfied identically.
\section{Summary and discussion}
To summarize let us quote the main results of this paper:
\begin{itemize}
\item{The leading irrelevant perturbation (\ref{H_int}), which controls the 
deviation of critical lattice Ising model with periodic boundary 
conditions from its continuous CFT analog is identified. The relation 
(\ref{g_value}) between anisotropy parameter and the coupling constant is 
established}
\item{
Next to leading $\sim 1/N^2$ corrections to the spectrum are calculated independently from 
lattice theory and from the perturbed CFT for several classes of states always 
finding exact agreement (subsections \ref{3.2} - \ref{3.4}).
}
\item { It is expected that to mimic higher order corrections, one should add 
	to unperturbed Hamiltonian terms, proportional to the higher integrals of motion in Ising CFT. In this way the integrable structure of the lattice theory 
	gets related to the integrable structure of CFT.
}
\end{itemize}
Besides square lattice, 2d Ising model is exactly solvable also on a number of other lattices with different symmetries. It would be interesting to identify the corresponding perturbing fields also in these cases. 
Clearly, it is expected that the spin of a perturbing field should be 
consistent with the order of discrete rotations allowed by respective lattice symmetry.   

It must be also possible to extend our analysis to the entire spectrum 
and higher orders in $1/N$ expansion, but this is left for future work.

\vspace{1cm}
\section*{Acknowledgments}

I am grateful to my supervisor Prof. Nikolay Izmailyan and Prof. 
Rubik Poghossian for introducing me into 
this field and for stimulating discussions.

This work was partially supported by the Armenian State Committee of Science
in the framework of the research project 18T-1C113.
\newpage
%\bibliography{references_PI}
\bibliographystyle{JHEP}
\providecommand{\href}[2]{#2}\begingroup\raggedright\endgroup

\end{document}